
\magnification=\magstep1

\def \w {\omega}

\def \tr{{\rm tr}}

\pageno=0
{\nopagenumbers
\line{\hfil HU-TFT-93-63}
\line{\hfil December, 1993}
\vskip 3 cm
\centerline {\bf Generalized Bethe Ansatz Equations for Hofstadter Problem}
\vskip 2 cm
\centerline {L.D. Faddeev$^{1,2,}$\footnote{$^\dagger$}
{Supported by the Russian Academy of Sciences and Academy of Finland},
R.M. Kashaev$^{1,}$\footnote{$^*$}
{On leave of absence from St. Petersburg Nuclear Physics
Institute, Gatchina, St. Petersburg 188350, Russia}$^{,\dagger}$}
\vskip 1 cm
\centerline {\it $^1$Research Institute for Theoretical Physics,}
\centerline {\it P.O.Box 9 (Siltavuorenpenger 20C), SF-00014 University
of Helsinki, Finland}
\vskip 0.5 cm
\centerline{\it $^2$St. Petersburg Branch of the Steklov Mathematical
Institute,}
\centerline{\it Fontanka 27, St. Petersburg 191011, Russia}
\vskip 2 cm
\line{\bf Abstract \hfil}

The problem of diagonalization of
the quantum mechanical Hamiltonian, governing dynamics of an
electron on a two-dimensional triangular or square
lattice in external uniform magnetic field, applied perpendicularly
to the
lattice plane, the flux through lattice cell, divided by the elementary
quantum flux, being rational number, is reduced to generalized Bethe
ansatz like equations on high genus algebraic curve. Our formulae for the
trigonometric case, where genus of the curve vanishes, contain as a
particular case recent result of Wiegmann and Zabrodin.
\vfil
\eject}

\beginsection{1. Introduction}

In this paper we consider the diagonalization problem of the
following Hamilton operator:
$$
{\cal H}=\mu(\alpha S+\alpha^{-1}S^{-1})
+\nu(\beta T+\beta^{-1}T^{-1})+\rho(\gamma U+\gamma^{-1}U^{-1}),\eqno(1.1)
$$
where unitary operators $S,T$,and $U$ satisfy Weyl commutation relations
$$
ST=\w TS,\quad TU=\w UT,\quad US=\w SU                \eqno(1.2)
$$
with $\w$ being a primitive $N$-th root of unity:
$$
\w=\exp(2\pi i M/N),\quad (M,N)=1                          \eqno(1.3)
$$
for some mutually prime integers $N>M\ge1$, complex parameters
$\alpha,\beta,\gamma$ have unit absolute values, while $\mu$, $\nu$ and
$\rho$ are real parameters. Phases of $\alpha,\beta,\gamma$ are
constrained to lay betweem 0 and $2\pi/N$:
$$
0\le{\rm arg}(\alpha),\quad{\rm arg}(\beta),\quad{\rm arg}(\gamma)\le2\pi/N.
                                                  \eqno(1.4)
$$
 Hamiltonian (1.1) appears in various problems of solid state physics.
With parameter $\rho$, taken to be zero, it governs the dynamics of an
electron on a two-dimensional square
lattice in the external uniform magnetic field, applied perpendicularly to the
lattice plane, phases of $\alpha$
and $\beta$ playing the role
of quasi-momentum, and the phase of $\w$, the role of magnetic flux
through lattice cell, divided by the elementary quantum flux.
The same problem on triangular lattice is
described by (1.1) with non-zero parameter $\rho$. Physical meanings of
parameters $\mu$, $\nu$ and $\rho$ are length scales along corresponding
directions on the lattice.

$N$-th powers of operators $S,T,U$ are central elements and we
fix them to be unity:
$$
S^N=T^N=U^N=1.                                        \eqno(1.5)
$$
{}From (1.2) and (1.5) one can deduce that operator
$$
C=STU                                                \eqno(1.6)
$$
is also central element, and its $N$-th power is 1 ($-1$) for odd (even) $N$:
$$
C^N=(-1)^{N-1}.                                              \eqno(1.7)
$$
If we fix $C$ by some number, satisfying (1.7), then operators $S,T,U$
have a unique $N$-dimensional irreducible representation.
For the problem under consideration only algebraic properties of these
operators, given by
(1.2), (1.5) are important. Different reducible realizations can affect
degeneracies of eigenvalues, but not the spectrum itself. We will use
this fact, working with $N^3$-dimensional reducible representation.

Exploration of the spectrum of (1.1), or its various particular cases,
is associated with many names, and especially Hofstadter, who at
$\rho=0$ made a
detailed numerical analysis of dependence of this spectrum  on integers
$M$ and $N$ for $N\le 50$ [H] . Shortened list of physical and
mathematical papers,
concerning this problem, contains also [W], [HK], [BKS], [CEY].

Last years the idea about possible
relevance to above problem of methods, used in quantization of
integrable models, attracted some attention. This idea has been materialized
in recent paper [WZ], where Wiegmann and
Zabrodin showed that in the case, equivalent to $\rho=0$, $\mu=-\nu=1$, and
$\alpha^N=\beta^{N}=1$, the spectrum of (1.1) can
be represented in terms of solutions  of Bethe ansatz (BA) type algebraic
equations.

In this paper, using more sophisticated methods of integrable models, we
show that BA like equations can be obtained also for general
Hamiltonian (1.1). They are similar to the equations, found in [BS]
for the case of the chiral Potts model.
Rapidity variables, satisfying generalized BA
equations, live on high genus algebraic curve. The
rational limit, where the genus vanishes, contains also the case,
considered in [WZ].
We hope, that our result will be useful for further investigation
of Hamiltonian (1.1), especially in the limit of large $N$.

In Section 2 we represent Hamiltonian (1.1) as a part of
three-site transfer matrix, constructed through elementary $L$-operator,
intertwined by the six-vertex $R$-matrix. In Section 3 we
calculate Baxter's vector, which enables to derive functional equation,
determining in principle the spectrum of Hamiltonian (1.1) in terms of
solutions of BA like equations on high genus algebraic curve. Section 4
contains specialization to the case with $\rho=0$ in (1.1). In Section 5
a detailed investigation of the trigonometric case,
corresponding to genus zero curve, is performed.

\beginsection{2. Transfer Matrix}

In this section we introduce a simple $L$-operator, intertwined by the
six-vertex $R$-matrix, and define transfer matrix, containing
Hamiltonian (1.1).

  Consider two operators, $X$ and $Y$,
satisfying Weyl permutation relation, and with unit $N$-th powers:
$$
XY=\w YX,\quad X^N=Y^N=1,                                    \eqno(2.1)
$$
with $\w$ from (1.3). These operators can be
realized as $N$-by-$N$ matrices with the following matrix elements:
$$
\langle m|X|n\rangle=\w^m\delta_{m,n},\quad
\langle m|Y|n\rangle=\delta_{m,n+1},                               \eqno(2.2)
$$
where indices $m$ and $n$ run over $N$ values $0,1,\ldots,N-1$ and are
considered $\pmod N$, so they are elements of $Z_N$.

Introduce 2-by-2 matrix, $L(x,h)$, with operator valued matrix
elements:
$$
L(x,h)=\pmatrix{aXY     &xbY\cr
              xcX&d\cr},                               \eqno(2.3)
$$
where $x$ and $h=(a,b,c,d)$ are some complex parameters. The $N$-dimensional
linear space, where operators $X$ and $Y$ act, will be refered to as
a ``quantum'' space while the two-dimensional space, where $L(x,h)$ acts
as 2-by-2 matrix, as an ``auxiliary'' one.

 Matrix (2.3), called $L$-operator, is intertwined by the six-vertex
$R$-matrix:
$$
R(x/y)L(x,h)\otimes L(y,h)=(1\otimes L(y,h))(L(x,h)\otimes1)R(x/y),\eqno(2.4)
$$
where $L(x,h)$ and $L(y,h)$ act independently in two different
auxiliary spaces and in one and the same quantum space, while $R(x/y)$ is
a matrix in the tensor product of the auxiliary spaces with numerical matrix
elements:
$$
R(x)=\pmatrix{x\w-x^{-1}&0&0&0\cr
                   0&\w(x-x^{-1})&\w-1&0\cr
                   0&\w-1&x-x^{-1}&0\cr
                   0&0&0&x\w-x^{-1}\cr}              .      \eqno(2.5)
$$
Note, that this $R$-matrix differs from the usual one in two diagonal elements.
Our choice enables us to escape square roots of $\w$ for the time being and
to consider both odd and even $N$ simultaneously.
Matrix (2.3) has been used in [BBP] for derivation of functional equations for
the chiral Potts model. It can be extracted also by specialization of
parameters
in $L$-operators, written in [BKMS] and [T].

Now introduce the following transfer matrix, acting in the tensor
product of three quantum spaces:
$$
{\cal T}(x)=\tr(L(x,h_0)\otimes L(x,h_1)\otimes L(x,h_2)),
\eqno(2.6)
$$
where the matrix products and trace are performed in the auxiliary space,
 and the indexed $h$'s mean that
three different sets
of parameters $h_i=(a_i,b_i,c_i,d_i)$ are taken.
To write down explicitly ${\cal T}(x)$, introduce special notations for
certain combinations of parameters $h_i$:
$$
e_i=b_{i-1}d_ic_{i+1},\quad f_i=c_{i-1}a_ib_{i+1},\quad i=0,1,2\pmod3,
                                                        \eqno(2.7)
$$
where index $i$ is considered as an element of $Z_3$. Then, for ${\cal T}(x)$
we have the following explicit expression:
$$
{\cal T}(x)={\cal T}_0+x^2{\cal T}_2,                         \eqno(2.8)
$$
where
$$
{\cal T}_0=\w a_0a_1a_2 C+d_0d_1d_2,  \eqno(2.9)
$$
and
$$
{\cal T}_2=e_2 S+f_2CS^{-1}+e_1T+f_1 CT^{-1}+e_0 U+f_0 CU^{-1},  \eqno(2.10)
$$
with
$$
S=X\otimes Y\otimes 1,\quad T=Y\otimes 1\otimes X,\quad
U=1\otimes X\otimes Y,               \eqno(2.11)
$$
$$
C=\w^{-1}XY\otimes XY\otimes XY.     \eqno(2.12)
$$
Operators $S,T,U$ and $C$, defined in (2.11) and (2.12),
satisfy algebraic relations (1.2), (1.5), and (1.6), operator
$C$ being commutative
with others (so it can be considered as a number). Thus,
operators (1.1) and (2.10) can be identified:
$$
{\cal T}_2={\cal H},     \eqno(2.13a)
$$
where
$$
\alpha=e_2^{1/2}(f_2C)^{-1/2},\quad
\beta=e_1^{1/2}(f_1C)^{-1/2},\quad
\gamma=e_0^{1/2}(f_0C)^{-1/2},         \eqno(2.13b)
$$
$$
\mu=(e_2f_2C)^{1/2},\quad
\nu=(e_1f_1C)^{1/2},\quad \rho=(e_0f_0C)^{1/2}.   \eqno(2.13c)
$$

To conclude the section note, that one and the same transfer matrix
${\cal T}(x)$ can be represented by (2.6) with different $L$'s:
$$
{\cal T}(x)=\tr(\tilde L_0(x,h_0)\otimes\tilde L_1(x,h_1)\otimes
\tilde L_2(x,h_2)),                                      \eqno(2.14)
$$
where $\tilde L$'s differ from $L$'s by a gauge transformation:
$$
\tilde L_i(x,h_i)=A_iL(x,h_i)A_{i+1}^{-1},\quad i\in Z_3      \eqno(2.15)
$$
with some invertible numerical 2-by-2 matrices $A_i$. In the next
section we will use this freedom to construct Baxter's vector.

\beginsection{3. Baxter's Vector}

Following [B], [FT], and [BS], let us turn to the calculation of Baxter's
vector for transfer matrix (2.6). For this let us use its representation
by (2.14) with matrices $A_i$, chosen as

$$
A_i=\pmatrix{1&\eta_i\cr
             1&\xi_i\cr}, \quad \eta_i=\xi_i-1,\quad i\in Z_3, \eqno(3.1)
$$
where $\xi_i$ are some complex parameters to be fixed later on.
Transformed $L$-operators $\tilde L_i(x,h_i)$ have the following explicit form:
$$
\tilde L_i(x,h_i)=\pmatrix{F(x,h_i;\eta_i,\xi_{i+1})
&-F(x,h_i;\eta_i,\eta_{i+1})\cr
F(x,h_i;\xi_i,\xi_{i+1})&-F(x,h_i;\xi_i,\eta_{i+1})\cr},\quad i\in Z_3,
          \eqno(3.2)
$$
where operator valued function $F$ is defined by
$$
F(x,h;\xi,\xi')=(\xi'aX-xb)Y+\xi(x\xi'cX-d).                    \eqno(3.3)
$$
Now let us try to fix parameters $\xi_i$ by demanding, that equations
$$
F(x,h_i;\xi_i,\xi_{i+1})|\phi_i\rangle=0,\quad i\in Z_3       \eqno(3.4)
$$
have non-zero solutions. Writing them in the basis (2.2), we get the
following recurrence relations for matrix elements of vectors
$|\phi_i\rangle$:
$$
{\langle m|\phi_i\rangle\over\langle m-1|\phi_i\rangle}=-\xi_i^{-1}
{\xi_{i+1}a_i\w^m-xb_i\over x\xi_{i+1}c_i\w^m-d_i},
\quad i\in Z_3.                                         \eqno(3.5)
$$
Periodicity conditions impose algebraic relations of high order on
$\xi$'s:
$$
\xi^N_i=(-1)^N{a_i^N\xi_{i+1}^N-b_i^Nx^N\over
c_i^N(x\xi_{i+1})^N-d_i^N},\quad i\in Z_3.               \eqno(3.6)
$$
Considering $\xi$'s and $x$ as variables, while keeping all $h$'s fixed,
we get three equations on four variables, which define an algebraic curve
$\Gamma$. Let $p$ be some point of
this curve, specified by a particular set of coordinates
$(x,\xi_i)$, satisfying (3.6). Then vectors $|\phi_i\rangle$
depend on this point, so from now on we will use another notation for
them:
$$
|\phi_i\rangle=|p\rangle_i,\quad p\in\Gamma.                   \eqno(3.7a)
$$
Besides, choose the following
normalization:
$$
\langle0|p\rangle_i=1,\quad i\in Z_3.                  \eqno(3.7b)
$$
To proceed further, define two automorphisms of our curve,
$\tau_\pm$, which act on coordinates $x$ and $\xi_i$ as follows:
$$
\tau_\pm:\Gamma\to\Gamma,\quad
\tau_\pm^*x=\w^{\pm1/2}x,\quad \tau_\pm^*\xi_i=\w^{-1/2}\xi_i. \eqno(3.8)
$$
Using this definition, one can easily show that
$$
F(x,h_i;\xi_i-1,\xi_{i+1})|p\rangle_i
=-|\tau_-p\rangle_i (x\xi_{i+1}c_i-d_i),           \eqno(3.9a)
$$
and
$$
F(x,h_i;\xi_i,\xi_{i+1}-1)|p\rangle_i
=-|\tau_+p\rangle_i \xi_i(a_id_i-x^2b_ic_i)/(\xi_{i+1}a_i-xb_i).\eqno(3.9b)
$$
Denoting
$$
|p\rangle=|p\rangle_0\otimes|p\rangle_1\otimes|p\rangle_2,
                                                 \eqno(3.10)
$$
and using (2.14), (3.2), (3.4), (3.9) together with multiplication properties
of triangular matrices, we come to the relation:
$$
{\cal T}(x)|p\rangle=|\tau_-p\rangle\Delta_-(p)+
|\tau_+p\rangle\Delta_+(p),                       \eqno(3.11)
$$
where
$$
\Delta_-(p)=\prod_{i\in Z_3}(d_i-x\xi_{i+1}c_i),\eqno(3.12a)
$$
$$
\Delta_+(p)=\prod_{i\in Z_3}\xi_i(a_id_i-x^2b_ic_i)/(\xi_{i+1}a_i-xb_i).
                            \eqno(3.12b)
$$

To get the functional equation from (3.11) we just multiply it from the
left by the eigenvector $\langle\varphi|$ of ${\cal T}(x)$, corresponding to
some eigenvalue $\Lambda(x)$. Vector $\langle\varphi|$ evidently
does not depend on $p$, so we obtain eventually
the scalar relation
$$
\Lambda(x)Q(p)=Q(\tau_-p)\Delta_-(p)
+Q(\tau_+p)\Delta_+(p), \eqno(3.13)
$$
where
$$
Q(p)=\langle\varphi|p\rangle         \eqno(3.14)
$$
is a function on our algebraic curve with known poles (they can be
extracted from recurrence relations (3.5)), while its zeros,
the number of them being equal to that of poles,
are determined from the generalized Bethe ansatz equations, obtained
from (3.13) by taking $p$ as various zeros of $Q(p)$:
$$
{Q(\tau_-p_k)\over
Q(\tau_+p_k)}=-{\Delta_+(p_k)\over\Delta_-(p_k)},\quad
Q(p_k)=0,\quad k=1,\ldots,\#({\rm poles}).                      \eqno(3.15)
$$
Relations (3.13) -- (3.15) at this stage only in principle
solve the diagonalization problem of (1.1), since eqs. (3.15) are too
complicated to work with. The main problem, of course, is that of
the suitable choice of coordinates on $\Gamma$. Unfortunately,
uniformization of algebraic curves
is still unsolved in general, long standing problem of mathematics. Our hope,
however, is that $\Gamma$ has a very special
structure, which we believe should admit particular approach to it.
 Anyway, we think that the analytic structure of $\Gamma$ deserves a further
study.

\beginsection{4. Hofstadter Hamiltonian}

Here we specify results of Section 3 to the case with $\rho=0$ in (1.1),
which corresponds to
$$
a_0=d_0=0,\quad b_0=c_0=1.                                    \eqno(4.1)
$$
The algebraic curve $\Gamma$, defined by (3.6), reduces to a disjoint
set of $N$ copies of another curve, $\Gamma_0$, defined by two high order
equations on three variables $\xi_0,\xi_2$ and $x$:
$$
\xi^{-N}_0=-{a_1^N\xi_{2}^N-b_1^Nx^N\over
c_1^N(x\xi_{2})^N-d_1^N},\quad
\xi^N_2=(-1)^N{a_2^N\xi_{0}^N-b_2^Nx^N\over
c_2^N(x\xi_{0})^N-d_2^N},               \eqno(4.2)
$$
while $s$-th copy in the set is specified by the value of $\xi_1$:
$$
\xi_1=\w^{s-1/2}/\xi_0,\quad s\in Z_N                               \eqno(4.3)
$$
with $-\w^{1/2}$ being chosen as $N$-th root of minus unity:
$$
(-\w^{1/2})^N=-1.                                        \eqno(4.4)
$$
Now Baxter's vector $|p\rangle$ as well as functions $\Delta_\pm(p)$ acquire
index $s$:
$$
|p\rangle\to|p,s\rangle,\quad \Delta_\pm(p)\to\Delta_\pm(p,s),\quad s\in
Z_N  \eqno(4.5)
$$
so relation (3.11) now reads
$$
{\cal T}(x)|p,s\rangle=|\tau_-p,s-1\rangle\Delta_-(p,s)+
|\tau_+p,s-1\rangle\Delta_+(p,s),         \eqno(4.6)
$$
where automorphisms $\tau_\pm$ of $\Gamma_0$ are defined by (3.8)
with index $i$ being restricted to only two values, $0$ and $2$.
 Taking into account (4.3) together with (3.12), we have explicitly:
$$
\Delta_\pm(p,s)=\w^s\Delta_\pm(p,0).             \eqno(4.7)
$$
 Noting that
$$
\w^s=\Phi(s)\Phi(1)/\Phi(s-1),\quad \Phi(s)=\w^{s(s+N)/2},  \eqno(4.8)
$$
then multiplying (4.6) by $\w^{st}/\Phi(s)$ and summing over $s$, we get
Baxter's relation for the case of Hofstadter Hamiltonian:
$$
{\cal T}(x)|p,t\rangle'=|\tau_-p,t\rangle'\Delta_-(p,t)\Phi(1)+
|\tau_+p,t\rangle'\Delta_+(p,t)\Phi(1),\quad t\in Z_N         \eqno(4.9)
$$
where
$$
|p,t\rangle'=\sum_{s\in Z_N}|p,s\rangle\w^{st}/\Phi(s),       \eqno(4.10)
$$
and we used (4.7). Repeating arguments from the end of Section 3, one can get
direct counterparts of relations (3.13) -- (3.15).

\beginsection{5. Rational Limit}

In this section we consider rational limit of formulae from Section 3.

Let us restrict parameters $h_i$, $i\in Z_3$, by
$$
a_i=q^{-1}d_i,\quad b_i=q^{-1}c_i,\quad i\in Z_3.          \eqno(5.1)
$$
Everywhere in this section $q=\w^{1/2}$ satisfies (4.4). Then, relations (3.6)
become very simple:
$$
\xi_i^N=\xi^N,\quad i\in Z_3,\quad \xi^{2N}=1.                   \eqno(5.2)
$$
Here variable $x$ does not enter, so, we have just several copies of genus
zero curve, spanned by coordinate $x$, while all $\xi$'s are fixed up to
$N$-th roots of unity. Let us put $\xi=q^l$ for some $l=0,\ldots,2N-1$ and
choose
$$
\xi_i=\xi,\quad i\in Z_3,                  \eqno(5.3)
$$
then, point $p$ of $\Gamma$ can be identified with pair $(x,l)$, and
(3.11) becomes
$$
{\cal T}(x)|x,l\rangle=|xq^{-1},l-1\rangle\Delta_-(x,l)+
|xq,l-1\rangle\Delta_+(x,l),\quad l=0,\ldots,2N-1,  \eqno(5.4)
$$
where
$$
\Delta_-(x,l)=\prod_{i\in Z_3}(d_i-c_ixq^l),          \eqno(5.5a)
$$
$$
\Delta_+(x,l)=\prod_{i\in Z_3}(d_i^2-c_i^2x^2)/(d_i-c_ixq^{-l}). \eqno(5.5b)
$$
Define function
$$
f(x,l)=\prod_{i\in Z_3}\prod_{j=0}^{[l/2]}
(d_i-xc_iq^{-l}\w^j)/(d_i-xc_iq^l\w^{-j}),\quad l=0,\ldots,2N-1,  \eqno(5.6a)
$$
where
$$
[l/2]=\cases{l/2,&$l=0\pmod2;$\cr
             (l-1)/2,&$l=1\pmod2,$\cr}  \eqno(5.6b)
$$
and for any $m\in Z_N$ introduce new vectors
$$
\eqalign{
|x,m\rangle_e&=\sum_{n=0}^{N-1}|x,2n\rangle f(x,2n)\w^{mn},\cr
|x,m\rangle_o&=\sum_{n=0}^{N-1}|x,2n+1\rangle f(x,2n+1)\w^{mn}.\cr}\eqno(5.7)
$$
Combining them into 2-component row vectors
$$
|x,m\rangle\rangle=\pmatrix{|x,m\rangle_e&
                    |x,m\rangle_o\cr},\eqno(5.8)
$$
one can rewrite (5.4) in a matrix form:
$$
{\cal T}(x)|x,m\rangle\rangle=
|xq^{-1},m\rangle\rangle D_-(x,m)
+|xq,m\rangle\rangle D_+(x,m),     \eqno(5.9)
$$
where
$$
D_\pm(x,m)=\pmatrix{0&\Delta_\pm(x,-1)\cr
         \w^m\Delta_\pm(x,0)&0\cr}.\eqno(5.10)
$$
Relations (5.9) do not change their form under gauge transformations:
$$
\eqalign{
|x,m\rangle\rangle&\to |x,m\rangle\rangle U(x,m),\cr
D_\pm(x,m)&\to U(xq^{\pm1},m)^{-1}D_\pm(x,m)U(x,m)\cr} \eqno(5.11)
$$
for any invertible 2-by-2 matrix $U(x,m)$. In the case of arbitrary $N$
 and without further restrictions
on parameters, there is no a gauge, where matrices
$D_\pm(x,m)$ would be diagonal, and rational in $x$, so, from now on let
us work with only odd $N$:
$$
N=2P+1,\quad P\ge 1.                                             \eqno(5.12)
$$
In this case $q=\w^{1/2}$, satisfying (4.4), can be represented as an
integer power of $\w$:
$$
q=\w^{1/2}=\w^{P+1}.            \eqno(5.13)
$$
Define function
$$
u(x)=\prod_{i\in Z_3}\prod_{j=0}^P(d_i-xc_i\w^j), \eqno(5.14)
$$
and choose gauge transformation matrix in (5.11) of the form
$$
U(x,m)=\pmatrix{u(qx)&q^{-m}u(x)\cr
                -q^mu(qx)&u(x)\cr}.    \eqno(5.15)
$$
Then matrices $D_\pm(x,m)$ become diagonal:
$$
D_\pm(x,m)=q^m\psi(\pm xq^{\pm1/2})
\pmatrix{1&0\cr0&-1\cr},                          \eqno(5.16)
$$
where
$$
\psi(x)=\prod_{i\in Z_3}(d_i+ xc_iq^{-1/2}).      \eqno(5.17)
$$
Now multiplying (5.9) from the left by the eigenvector
$\langle\varphi|$ of ${\cal T}(x)$, corresponding to some eigenvalue
$\Lambda(x)$, we get two scalar relations:
$$
\pm q^{-m}\Lambda(x)Q_\pm(x)=\psi(-xq^{-1/2})Q_\pm(xq^{-1})
+\psi(xq^{1/2})Q_\pm(xq),                       \eqno(5.18)
$$
where
$$
\pmatrix{Q_+(x)&Q_-(x)}=\langle\varphi|x,m\rangle\rangle.  \eqno(5.19)
$$
{}From the structure of Baxter's vector, given by (3.5), it follows that
$Q_\pm(x)$ is a polynomial in $x$, and sending $x$ to zero and
infinity, taking into account (2.8) and (2.9), we
conclude, that
$$
Q_-(x)=0,\quad Q_+(x)=x^PQ(x),\quad {\rm deg}Q(x)=2P,\quad m=P,
                               \eqno(5.20)
$$
while central element $C$ in (1.6) equals to unity.
For convenience choose the following normalization for the parameters
$c_0,c_1,c_2$\footnote{$^\star$}{This can be done by rescaling variable $x$}:
$$
c_0c_1c_2=\w.                                              \eqno(5.21)
$$
Then, function $\psi(x)$ reads
$$
\psi(x)=q^{1/2}(x+\mu)(x+\nu)(x+\rho),                  \eqno(5.22)
$$
where $\mu,\nu,\rho$ are the same as those in (1.1). As for the parameters
$\alpha,\beta,\gamma$, they are equal to $q^{1/2}$. Using (5.20), we rewrite
(5.18) as
$$
\Lambda(x)Q(x)=\psi(-xq^{-1/2})Q(xq^{-1})
+q^{-1}\psi(xq^{1/2})Q(xq),\quad {\rm deg}Q(x)=2P,     \eqno(5.23)
$$
$\Lambda(x)$ being
$$
\Lambda(x)=\mu\nu\rho(q^{1/2}+q^{-1/2})+x^2E,      \eqno(5.24)
$$
where $E$ is an eigenvalue of Hamiltonian (1.1).
Let $Q(x)$ has the following decomposition:
$$
Q(x)=\prod_{m=1}^{2P}(x-1/z_m)                        \eqno(5.25)
$$
for some $z$'s, which should satisfy BA equations (3.15):
$$
q^{-1/2}{(\mu z_l+q^{1/2})(\nu z_l+q^{1/2})(\rho z_l+q^{1/2})
\over(q^{1/2}\mu z_l-1)(q^{1/2}\nu z_l-1)(q^{1/2}\rho z_l-1)}=
\prod_{m=1,m\ne l}^{2P}{qz_l-z_m\over z_l-qz_m},\quad l=1,\ldots,2P.\eqno(5.26)
$$
Differentiating (5.23) twice with
respect to $x$ at $x=0$, we get the expression for $E$ entirely in
terms of $z$'s:
$$
\eqalign{
E=\mu\nu\rho&(q-q^{-1})(q^{1/2}-q^{-1/2})
\sum_{1\le m<n\le2P}z_mz_n\cr
-&(q-q^{-1})(\mu\nu+\nu\rho+\rho\mu)\sum_{m=1}^{2P}z_m
+(q^{1/2}+q^{-1/2})(\mu+\nu+\rho).\cr}  \eqno(5.27)
$$
Formulae (5.26) and (5.27) at $\rho=0$, $\mu=-\nu=1$ reproduce correspondingly
formulae
(6) and (5) of paper [WZ], if we identify our $z_m$ with their $iz_m$.
\footnote{$^{\star\star}$}{In fact our eqs. (5.26) differ from (6) of [WZ] by a
sign. We checked our equations by direct diagonalization
at $N=3$, so we think that there is a misprint in [WZ].}

It is worth to note that in the case of odd $N$ one can start from the very
beginning with another $L$-operator:
$$
L(x,h)=\pmatrix{aV     &xbW\cr
              xcW^{-1}&dV^{-1}\cr},                               \eqno(5.28)
$$
where operators $V$ and $W$ satisfy
$$
VW=qWV,\quad V^N=W^N=1.                                   \eqno(5.29)
$$
$L$-operators (5.28) and (2.3) are connected in a simple way. If we
multiply (5.28) from the right by $V$, and replace parameter $a$ by $qa$, then
we get (2.3) with
$$
X=W^{-1}V,\quad Y=WV.                                        \eqno(5.30)
$$
$L$-operator (5.28) can be extracted as a particular case from the
$L$-operator, considered by Bazhanov and Stroganov in [BS].
The case, where $a=d$ and $b=-c$, is related to the massless quantum
Sine-Gordon model on the lattice [G], [V], [KT].

Using $L$-operator (5.28), we can repeat the procedure developed in
Sections 2 and 3. But automorphisms $\tau_\pm$ in this case act on variables
$\xi_i$
and $x$ in another way, namely, they do not change $\xi$'s at all. So,
in the rational limit we are not forced to use simultaneously several
disjoint copies of the zero genus curve.
Particularly, (5.23) and (5.26) arise as a direct specialization of
general formulae, counterparts of (3.13) and (3.15), without intermediate
transformations, which we made in this section. Nevertheless, the
approach which we have presented in detail, enables us to consider both even
and
odd $N$ on equal footing.
This explains why we prefer to work with $L$-operator (2.3).

\beginsection{Summary}

The main result of this paper consists in establishing the relation
between the generalized Hofstadter Hamiltonian (1.1) and inhomogeneous
$XXZ$ chain of three sites. The available methods in lattice integrable
models lead to the generalized Baxter relation (3.11), as well as Bethe
ansatz (BA) type equations (3.15) on high genus algebraic curve.
In particular rational case we have got BA like equations (5.26)
together with formula (5.27) for eigenvalues of the Hamiltonian (1.1).

To use Baxter's  relation in general case, one has to develop analysis on
the high genus algebraic curve. Apparently, one needs more information
about automorphic functions, connected with the curve.

The tantalizing opportunity also is to
try to apply the known simplification properties of BA equations when
the number of variables, satisfying these equations, tends to infinity.
In our case this corresponds to $N\to\infty$. However, at the moment it is not
clear, how to utilize this possibility.

\beginsection{Acknowledgements}

One of the authors (L.D.F.) got the idea of a possible connection of
the lattice Sine-Gordon chain to Hofstadter problem during his encounter
with R. Seiler in Berlin in February 1993. Then the talk of P.B.
Wiegmann at the Helsinki conference in September 1993 was very
stimulating for both of us. We express our gratitude to R. Seiler and
P.B. Weigmann for the most stimulating and helpful discussions.

\beginsection{References}

\item{[B]} R.J. Baxter, {\it Exactly solved models in statistical
mechanics}, Academic Press, New York, 1982

\item{[BBP]} V.V. Bazhanov, R.J. Baxter, J.H.H. Perk, J. Mod. Phys. B{\bf4}
(1990) 803

\item{[BKMS]} V.V. Bazhanov, R.M. Kashaev, V.V. Mangazeev,Yu.G.
Stroganov, Commun. Math. Phys. {\bf138} (1991) 393

\item{[BKS]} J. Bellissard, C. Kreft, R. Seiler, J. Phys. A: Math. Gen.
{\bf24} (1991) 2329

\item{[BS]} V.V. Bazhanov, Yu.G. Stroganov, J. Stat. Phys. {\bf59}
(1990) 333

\item{[CEY]} M-D. Choi, G.A. Elliott, N. Yuri, Invent. Math. {\bf99}
(1990) 225

\item{[FT]} L.D. Faddeev, L.A. Takhtajan, Usp. Mat. Nauk, {\bf34} (1979)
13 (in Russian)

\item{[G]} J.L. Gervais, Phys. Lett. B{\bf160} (1985) 279

\item{[H]} D.R. Hofstadter,
 Phys. Rev. B {\bf 14 } (1976) 2239

\item{[HK]} H. Hiramoto, M. Kohmoto, Int. J. Mod. Phys. B {\bf 6}
(1992) 281

\item{[KT]} V.E. Korepin, V.O. Tarasov, Remark on Australian National
University Workshop on Yang-Baxter equation, Canberra, 1989
(unpublished)

\item{[T]} V.O. Tarasov, Int. J. Mod. Phys. A {\bf7}, Suppl. 1B
(1992) 963

\item{[V]} A.Yu. Volkov, Theor. Math. Phys. {\bf74} (1988) 135

\item{[W]} G.H. Wannier, Phys. Status Solidi {\bf 88} (1978) 757

\item{[WZ]} P.B. Wiegmann and A.V. Zabrodin, {\it Bethe-Ansatz Approach to
Hofstadter Problem. Quantum Group and Magnetic Translations}, Preprint
LPTENS-93/34
\item{} See also in: L.Beaulieu, V. Dotsenko, V. Kazakov, P. Windey
(eds.) New developments in string theory, conformal models and
topological field theory. Proceedings, Cargese 1993. Plenum Press; to appear
\bye